# THE IMPACT OF SURROUNDING ROAD OBJECTS AND CONDITIONS ON DRIVERS ABRUPT HEART RATE CHANGES

Arash Tavakoli, and Arsalan Heydarian

Recent studies have pointed out the importance of mitigating drivers' stress and negative emotions. These studies show that certain road objects such as big vehicles might be associated with higher stress levels based on drivers' subjective stress measures. Additionally, research shows strong correlations between drivers' stress levels and increased heart rate (HR). In this paper, based on a naturalistic multimodal driving dataset, we analyze the visual scenes of driving in the vicinity of abrupt increases in drivers' HR for the presence of certain stress-inducing road objects. We show that the probability of the presence of such objects increases when becoming closer to the abrupt increase in drivers' HR. Additionally, we show that drivers' facial engagement changes significantly in the vicinity of abrupt increases in HR. Our results lay the ground for a human-centered driving experience by detecting and mitigating drivers' stress levels in the wild.

## INTRODUCTION

One of the key elements of human-vehicle interaction is the prediction and analysis of drivers' states and behaviors, such as stress, anxiety, and negative emotions. Detecting and mitigating drivers' stress levels and negative emotions on the road are of high importance for decreasing accident rates as well as providing a human-centered experience in driving (Balters et al., 2021; Chung et al., 2019). Additionally, research shows that driver-state changes are associated with certain environmental events. For instance, recent research suggests that driver stress level and anxiety are correlated with specific events, such as the presence of passengers, big vehicles, pedestrians, and intersections (Bustos et al., 2021; Zepf et al., 2019).

Integrating the above information into the human-vehicle interaction development can help mitigate the adverse effects of driving on the users and provide a human-centric driving experience. However, methods of data collection other than subjective self-reports are required to understand the real-time reaction of drivers to these environmental attributes (Balters et al., 2021). In this regard, studies have already shown the utility of drivers' psychophysiology in capturing drivers' reactions to environmental attributes. For instance, (Tavakoli, Kumar, Guo, et al., 2021) through manually annotating the visual scene showed that environmental changes such as the presence of lead vehicles are associated with an abrupt increase in drivers' HR.

Literature suggests that stress is the process in which the demand of a certain situation is perceived to be more than the available resources (Francis, 2018). The perceived demand can be defined based on the overall situation, including the previous experiences, internal body sensations, and external stimuli (Francis, 2018). The experienced stress may be accompanied by changes in physiology, such as increases in HR. Thus, it is possible that abrupt changes in drivers' HR are preceded by the presence of stress-inducing road objects (e.g., passing through an intersection or the presence of a big vehicle) (Dittrich, 2021).

In this paper, we use a naturalistic driving dataset, to analyze the probability of the presence of certain objects on the road before an abrupt increase in drivers' HR. In this regard, we use an off-the-shelf object detection algorithm to detect road objects.

We detect abrupt increases in drivers' HR using the Bayesian change point detection method. Coupling the results of object detection with change points in HR, we show that the probability of the presence of certain categories of road objects increases in the vicinity of abrupt increases in drivers' HR. Additionally, by using drivers' facial expressions, we show that facial engagement (depicting the emotional expressivity) also increases significantly in the vicinity of change points in HR. This paper builds the initial steps towards multimodal human-vehicle interaction models that can provide a potential explanation for sudden changes in driver's HR.

## BACKGROUND

Human biomarkers were used in the literature for detecting stress levels (Chesnut et al., 2021). In this regard, human HR has received special attention as HR can be collected using conventional wearable devices and have the potential to be used in the wild. Wearable devices often use photoplethysmogram (PPG) technology, which is based on using infrared to detect the blood volume pressure in the veins on the wrist (Tavakoli, Kumar, Boukhechba, et al., 2021). This is then used to estimate the HR and a set of heart rate variability (HRV) features. Using conventional wearable devices, studies showed that an increase in HR is correlated with an increase in the levels of certain states such as anxiety, stress, and anger (Kim et al., 2018). Changes in human biomarkers are also accompanied by variations in facial expressions while experiencing stress. For instance, (Giannakakis et al., 2017) showed that facial cues that are visible during a stress event are distinct and can be detected through machine learning methods.

Within the driving research area, studies have provided significant insights into the causation behind drivers' stress and negative emotions. These studies were mostly conducted in an on-road controlled fashion or within a driving simulator framework. In this regard, (Zepf et al., 2019) monitored 33 drivers during a 50 minutes naturalistic on-road controlled study through a combination of cameras, sensors, and self-reports through voice. Through analyzing 531 self-reports, the authors found that the four main types of emotion triggers include traffic & driving task, environment, HCI & navigation, and vehicle and equipment. Their study mentions different detailed categories such as traffic lights, road design, the behavior of others, weather, and building and sites as reasons within the aforementioned larger categories for negative emotions of drivers. Another study performed by (Tavakoli, Kumar, Guo, et al., 2021) identified that different characteristics of the scene might be associated with abrupt increases in drivers' HR. In their study, authors found lead vehicles, intersections, being followed by another vehicle, and performing secondary tasks to be the most significant factors associated with abrupt changes in drivers' HR. Later a study by (Bustos et al., 2021) attempted to classify drivers' self-reported stress levels using the objects in the scene. In their study, by applying computer vision methods such as convolutional neural networks on the outside videos as input, authors achieved an accuracy of 72 % in the prediction of drivers' self-reported stress in an on-road controlled study. Their study also points out that objects such as traffic signs, cars, pedestrians, big vehicles, and riders are associated with medium to high subjective stress levels. Another recent study performed by (Dittrich, 2021) tracked the emotions of 34 drivers through self-reports on an on-road controlled study and found out that intersections are the hotspots for emotional triggers. Additionally, they found out that highways are associated with the least stress level as compared to other urban environments. This study also points out that the behavior of other road users, traffic lights, and navigation have a higher fraction of negative emotions as compared to the category of entertainment, which had a more positive emotion.

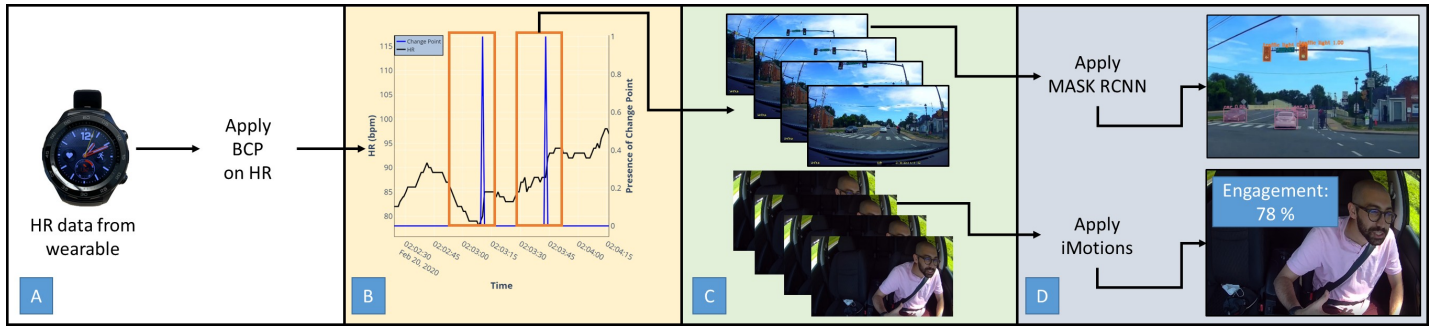

*Figure 1 The framework of the study.*

Research has shown the association between changes in drivers' physiological signals, facial expressions, and environmental attributes. A recent study through a naturalistic driving framework found out that drivers' have a lower heart rate when driving in clear versus rainy, highway versus city, and with passenger versus alone conditions (Tavakoli et al., 2020). Another study found out that drivers' physiological responses are correlated with the vehicle's kinematic, which is governed by the driver's behavior (Milardo et al., 2021). In their study, authors conducted a naturalistic study across 16 drivers. They concluded that drivers' HR decreases when the number of people in the car increases. Lastly, (Liu et al., 2021) proposed a framework to detect drivers' emotions using a combination of vehicle telemetry as well as the outdoor visual scene. Based on the facial expressions retrieved from the Affectiva software applied on 675 hours of driving data, authors were able to detect the emotions in a user-dependent and user-independent fashion with 70 % and 60 % accuracy, respectively.

The above literature taken together suggests certain objects on the road are correlated with higher subjective stress levels and negative emotions. Additionally, higher stress levels, anxiety, and negative emotions are also correlated with increases in HR relative to the resting condition. This allows us to hypothesize that:

- *Hypothesis 1:* Certain on-road objects (e.g., big vehicles) and infrastructures (e.g., intersections) might be correlated with abrupt increases in drivers' HR. In this regard, the probability of the presence of these on-road objects increases in the vicinity of abrupt increases in drivers' HR.
- *Hypothesis 2:* The overall presence of facial expressions might change in the vicinity of abrupt increases in drivers' HR showing possible signs of changes in drivers' state.

## METHOD

**Data Details**

The dataset for this study is provided by HARMONY, which is a naturalistic driving study framework that collects longitudinal driving data through cameras, smartwatches, and multiple APIs (Tavakoli, Kumar, Guo, et al., 2021). Every participant in this dataset is provided with a dual dash camera and an Android smartwatch, which monitors their HR, hand movements (i.e., IMU sensors), facial expressions, gaze direction, pose direction. Through these devices, vehicle's speed, location, as well as outdoor environmental videos are also collected. From this dataset, we utilize video streams and HR data from 15 participants (Figure 1- A). For each video recording, the synchronized HR file is generated based on the start and finish timestamp of each video. The wearable device collects HR with a 1 Hz frequency. The HR data is estimated from the PPG sensor on the watch, which measures the blood volume in the veins using infrared technology. The camera records both inside and outside conditions with 30 fps and a resolution of 1080P, and in 3-minute segments. A random subset of the data is chosen for our case study. The random subset of the data is from 15 participants (7 females and 8 males) who were between the ages of 21 to 33. From each participant, we used approximately 2 hours of driving data which includes around 200000 frames of videos.

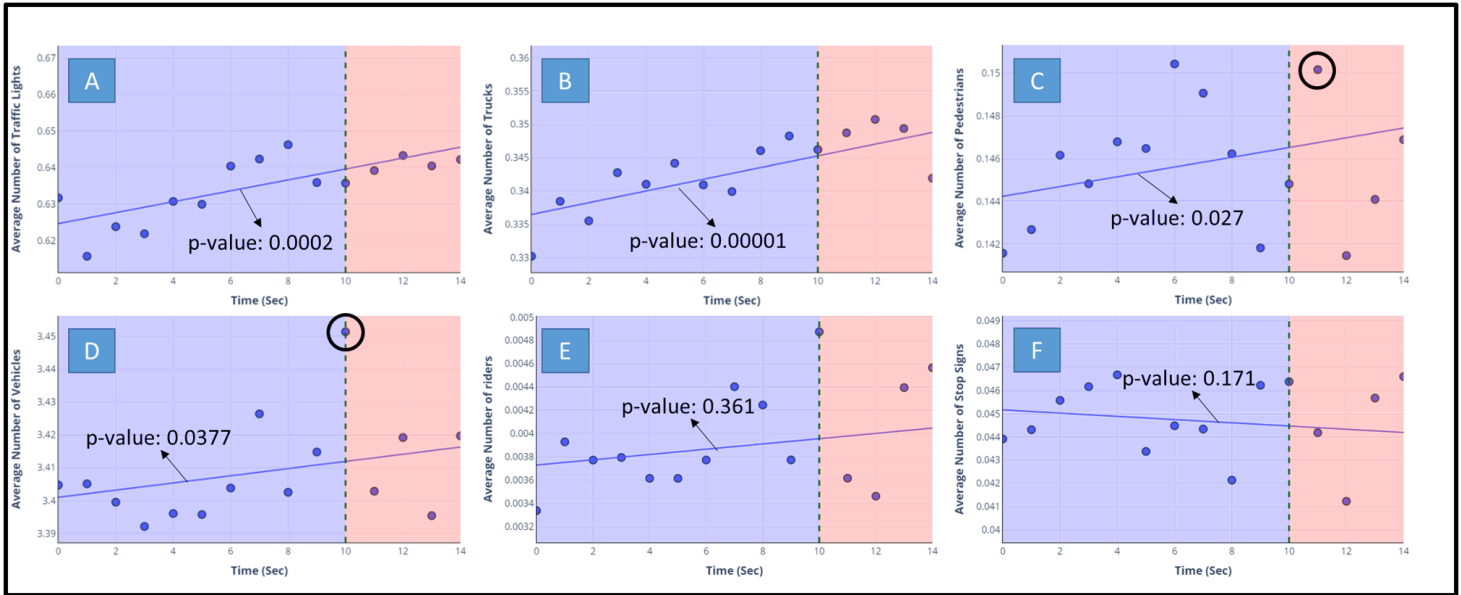

*Figure 2 The average number of each road object prior to change in HR. The dashed line shows the moment that change point is detected in drivers' HR. The blue and pink colors show the prior and after change point respectively. Note that the numbers are*

## Change Point Detection

As discussed previously, recent research suggests that an increase in HR might be correlated with an increase in stress levels and negative emotions. In order to detect the abrupt increases in drivers' HR, we take advantage of the change point detection method. Note that we are not interested in peaks in drivers' HR, which might be due to errors in HR estimation because of motion artifacts. However, we are interested in detecting a change in the overall distribution of HR. In this regard, we use a Bayesian Change Point (BCP) detector (Figure 1 - B). This approach allows for probabilistically detecting changes in the HR data points. Previous studies have pointed out the application of BCP in detecting changes in different fields such as health (Malladi et al., 2013). We leverage Barry and Hartigan's (Barry & Hartigan, 1993) Bayesian change point model for this analysis. In order to perform change point detection through this model, we use the bcp package written in R programming language (Erdman & Emerson, 2007). The BCP is applied to each participants' HR data, and locations with a high probability of change point (equal to or more than 0.8) are then extracted. For each change point, 10 seconds prior and 2 seconds after being extracted from the video streams (Figure 1 - C). The choice of the 2 seconds time window after the changepoint is solely for visualization purposes and does not affect our analysis as it has not been used in the modeling scheme. The extracted frames are then fed into (1) an object detection algorithm (for outside videos) (Figure 1 - D), and (2) a facial expression extraction software, Affectiva iMotion (for inside videos) (Figure 1 - D) as follows. This section is based on 6396 samples of change points across 15 participants in different road environments (i.e., highway and city).

## Road Object Detection

In order to detect different objects on the road, we take advantage of the current off-the-shelf object detection algorithms. Different object detection algorithms have been proposed and trained on different datasets in the past. For the purpose of this paper, we use the MASK R-CNN algorithm He, Gkioxari, Dollár, and Girshick (2017) trained on the Common Objects in Context (COCO) dataset Lin et al. (2014). MASK R-CNN is an object detection algorithm that works in two stages. First, it predicts the locations that might be associated with different objects within the field of view. Second, the algorithm predicts the class of the object.

We specifically chose the pretrained model on COCO as it has many road objects that were previously shown to be correlated with stress levels. We used the pre-trained models from the Abdulla (2017) repository. COCO includes car, pedestrian, rider (i.e., bicycle and motorcycle), truck, and traffic lights. A sample of the road objects detected by this method on our videos is shown in Figure 1 - D. This algorithm is applied to all the outside video frames that were retrieved from the previous section. The output of this algorithm is a JSON file, which includes the road objects bounding boxes as well as the confidence level. For each frame, the presence of the aforementioned road objects is saved into a cleaned CSV file. This information is then summarized for all of the instances of change points across participants and different driving situations. The detections with lower than 50 percent confidence by the algorithm were removed from the analysis. 20 percent of the samples were manually analyzed for validation. Additionally, if a detection did not persist for 80 percent of a 45 frame window, they were removed as false positives.

## Facial Expressions

In order to extract facial expressions, we use the Affectiva module on the iMotion software (McDuff et al., 2016). Previous studies have pointed out the utility of this software in extracting the level of showing facial expressions (i.e., engagement), the positivity or negativity of emotions (i.e., valence) emotions

(e.g., anger), and certain movements of facial muscles (e.g., smile) (Abdíc et al., 2016). we focused on the expressivity of emotions, which is referred to as engagement. Engagement is a value between 0 and 100, where zero depicts no evidence of showing emotion in the face of the participant and 100 shows very high visual facial expressions. After performing analysis with Affectiva, all the frames that did not have any detections were removed from the database. This can be due to the angle of the camera as well as lighting issues. Note that the number of such frames was less than 10 percent of the dataset.

## RESULTS

Figure 2 shows the average number of each road object 10 seconds prior to the change point in drivers' HR and 2 seconds after. In addition to the 5 objects, we also report the results for vehicle objects. The change point is shown with a dashed line in the figure. Based on findings of previous studies on self-reported stress levels and road objects, we focus on a set of five objects which were also available in the COCO dataset. Figure 2 - A shows the change in the number of traffic lights prior change point in HR. As depicted, the number of traffic lights increases in the frames that are closer to a change point in HR. We have then confirmed this visual inspection by applying a linear regression model. For each section of Figure 2, a linear regression model was applied predicting the average number of each road object based on the time prior to changepoint in HR. This is in line with previous studies depicting the increase in drivers' stress levels when reaching an intersection, as mentioned in the background section. The same trend is also happening within the number of trucks (Figure 2 - B). This is also confirmed through linear regression, showing that the time to the event is a significant predictor of the number of trucks with a p-value of less than 0.0001 and a high R2 of 0.7.

The average number of pedestrians and vehicles also increases in the vicinity of change point in HR. This is confirmed through a linear regression with p-values of less than 0.0001 for both pedestrians and vehicles. More interestingly, the number of pedestrians and vehicles has an abrupt increase ±1 second prior to the location of the change point (shown with black circles on Figure 2). This might indicate that momentarily changes in the number of vehicles and pedestrians are associated with the abrupt increases in HR as detected by the BCP. In other words, while the overall increase in the number of vehicles and pedestrians follows the same pattern as the number of trucks and intersections, the underlying reason for the change point, in this case, might be more related to the abrupt increase in the number rather than the overall gradual increase. This can also be further confirmed by the fact that the number of vehicles drops quickly after the change point (second 10-12 on Figure 2 - D). This might suggest that for different types of events, the research methodology should consider different timeframes prior to the change point for providing an explanation for each state change.

While the number of riders in Figure 2 - E suggests an overall increase when getting closer to the moment of change point, we did not find any significant result when regressing the number of riders with respect to time to change point. However, note that the number of riders also exhibits an abrupt increase when reaching the change point moment in HR. While Figure 2 - F might suggest a downward trend for the probability of stop signs prior to the change point, we did not find any significant results in the regression.

Additionally, We analyzed the proportion of the presence of each object within the 1-second vicinity of a change point in HR. This number is then normalized based on the total number of samples that are present. The results are depicted in Figure 3. Based on this figure, we observe that other vehicles, intersections, and trucks are among the top objects on the road that are accompanied by abrupt increases in drivers' HR.

Within the analyzed frames, we have compared the city and highway based on the average presence of each road object at the exact moment of HR change point. In this regard, we counted the number of each road object within all the frames associated with a change point in HR, which is shown in Table 1. As depicted, the average presence of most objects is higher in the city as compared to the highway. The table shows that in the highway environment, most of the HR change points are associated with the presence of trucks as compared to other road objects (e.g., traffic lights), whereas the city environment seems to have a myriad of objects being involved in the abrupt increases in drivers' HR.

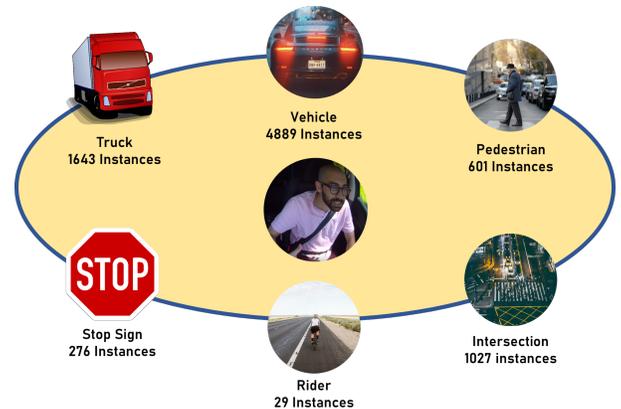

*Figure 3 The number of instances of each category of road objects that were accompanied by an abrupt increase in drivers' HR*

Table 1. The average presence of each road object within city and highway environment, within all of the instances of having a change point in HR.

| Driving Environment | Pedestrian | Vehicle | Truck | Traffic Light | Stop Sign | Rider |
|---|---|---|---|---|---|---|
| City | 0.152 | 0.803 | 0.109 | 0.308 | 0.064 | 0.006 |
| Highway | 0.050 | 0.742 | 0.377 | 0.046 | 0.027 | 0.003 |

We also analyzed the facial expressions of drivers, specifically the engagement output of Affectiva in the vicinity of change points in HR (Figure 4 - A). In this section, we considered a window of 10 seconds prior and after the change point (Figure 4). It is visible from the figure that the engagement has an abrupt increase at the time of the change point in HR. We have then confirmed this by applying the same change point detector to the engagement values. Note that here we detect two change points, one when the engagement increases (between seconds 10 and 11) and one after the engagement moves back to its regular values (around second 15).

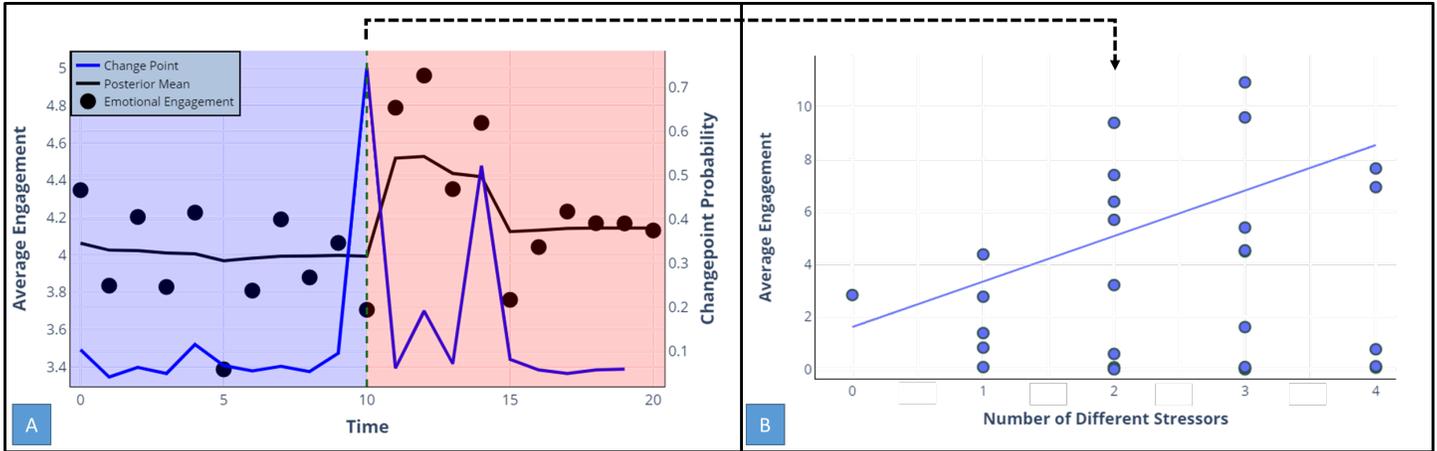

*Figure 4 (A) The change in drivers' emotional engagement over time. Note that the dashed green line shows the location of an abrupt increase in HR, which is also followed by an abrupt increase in drivers' emotional engagement.*

Lastly, we analyzed the average facial engagement for a cumulative sum of different stress-inducing road objects that were present simultaneously within the 10-second window of HR change points. For example, if traffic lights and trucks were present in the vicinity of an HR change point, we counted them as 2 objects. Figure 4 - B depicts the changes in average engagement for different numbers of stressors on the road. As depicted visually, the average engagement increases as the number of stressors increases. However, a linear regression did not produce significant results for this relationship, while this apparent increase stays an interesting starting point for further research.

## DISCUSSION

The goal of this paper is to analyze the presence of certain stress-inducing road objects detected automatically through computer vision in the vicinity of abrupt increases in drivers' HR retrieved from the Bayesian changepoint detector. While previous research was mostly centered around the relationship between road objects and subjective stress level (Bustos et al., 2021; Dittrich, 2021), we focused on drivers' HR as a proxy of stress level, which can be used in the wild. Our results indicated that the probability of the presence of certain road objects increases in the vicinity of change points in drivers' HR. Considering the fact that stress level is also correlated with abrupt increases in drivers' HR, this might indicate a causal relationship between environmental attributes and drivers' psychophysiology and state changes as changes in the environment are immediately being followed by changes in drivers' HR. Nevertheless, the association between road objects and drivers' psychophysiology has strong implications for the human-centered design of future vehicles, where mitigating drivers' negative emotions, stress levels, and anxiety is of interest. For instance, through personalized empathetic routing, transportation systems can provide a healthy driving experience (Tavakoli, Boukhechba, et al., 2021).

Previous research showed that highways are correlated with lower stress levels. This has been shown through both subjective stress levels (Bustos et al., 2021), as well as average HR being lower in highway environments (Tavakoli et al., 2020), and higher speeds (Liu et al., 2021) with lower acceleration. Our findings are also in line with previous research showing that highways exhibit fewer stressors as compared to the city environment. This might provide reasoning as to why people preferred highway driving in the past, both through subjective and objective measures.

In this research, we only focused on one object detection algorithm to contextualize the visual scene. In our analysis, we could not conclude that presence of stop signs increases as getting closer to HR change point. Our preliminary inspection of the samples that included stop signs shows that the computer vision algorithm has high false positives in our videos, mistaking other road signs with the stop sign. This was not the case for other objects used in this study. Our ongoing work is focused on developing a more accurate stop sign detection model.

Our results show that drivers' facial expressions measured through the engagement parameter of Affectiva also increase significantly in the vicinity of change points in HR. Additionally, we observe that the increase in engagement moves back to its prior value after the HR change point. This can indicate a sequence of state changes, starting from a perturbation in the environment (e.g., presence of a truck), leading to changes in drivers' HR, and facial expressions. The temporal and sequential aspects of the aforementioned changes are of importance and will be explored in future research. Note that here we only focused on the expressivity of emotions rather than the positivity or negativity of emotions. Future research will also consider other dimensions of emotions, such as valence.

While our preliminary analysis provides specific insights into the interaction between the road objects and drivers' HR, more research is required to understand the depth of the association between them as well as individual differences across participants. The future research is focused on multiple aspects. First, we will increase the number of instances of analysis for each road perturbation. Second, we will analyze the effect of each perturbation on each participant separately to understand the individual differences across people when facing road objects. Third, we will use more detailed machine learning algorithms to detect other road objects (e.g., guardrails) as well as enhance the detection for some of the categories in the current research (e.g., stop sign). Fourth, we will apply more detailed facial expression analysis on the drivers' facial videos to understand other dimensions of drivers' emotions around road objects.

## CONCLUSION

This study analyzed the association between certain road objects and abrupt increases in drivers' HR. Based on a naturalistic dataset, this study first retrieved the abrupt increases in drivers' HR using a Bayesian change point detector. Then by using computer vision applications, different road objects in the field of view were retrieved. The results suggest that the presence of certain road objects that were previously shown to be associated with an increase in drivers' subjective stress level, increases when getting closer to the HR change point location.